\documentclass[prl,aps,twocolumn,groupedaddress,showpacs,floatfix,preprintnumbers]{revtex4}
\usepackage{amsmath,amssymb,multirow,epsfig,bm}

\newcommand{\etal}{{\it et al.,\;}}
\newcommand{\avg}[1]{\langle #1 \rangle}
\newcommand{\eqn}[1]{Eq.~(\ref{#1})}
\newcommand{\fig}[1]{Fig.~\ref{#1}}
\newcommand{\apriori}{\textit{a~priori} }              
\newcommand{\pictsize}{0.92}
\newcommand{\beq}{\begin{equation}}
\newcommand{\eeq}{\end{equation}}
\newcommand{\bea}{\begin{eqnarray}}
\newcommand{\eea}{\end{eqnarray}}
\newcommand{\pcut}{p_{\text{max}}}  
\begin{document}
\preprint{LA-UR-12-20206/LA-UR-12-20207}
\preprint{NT@UW-12-04}

\title{Shear Viscosity of a Unitary Fermi Gas}

\author{Gabriel Wlaz\l{}owski,$^{1,2}$ Piotr Magierski,$^{1,2}$ and Joaqu\'{\i}n E. Drut$^3$}

\affiliation{$^1$Faculty of Physics, Warsaw University of Technology,
Ulica Koszykowa 75, 00-662 Warsaw, POLAND}
\affiliation{$^2$Department of Physics, University of Washington, Seattle,
Washington 98195--1560, USA}
\affiliation{$^3$Theoretical Division, Los Alamos National Laboratory, Los Alamos, New Mexico 87545--0001, USA}

\begin{abstract}
We present an {\it ab initio} determination of the shear viscosity $\eta$ of the unitary Fermi gas, based on
finite temperature quantum Monte Carlo calculations and the Kubo linear-response formalism.
We determine the temperature dependence of the shear viscosity-to-entropy density ratio $\eta/s$.
The minimum of $\eta/s$ appears to be located above the critical temperature for 
the superfluid-to-normal phase transition with the most probable value being $(\eta/s)^{}_{\textrm{min}}\approx0.2\,\hbar/k^{}_{B}$,
which is close the Kovtun-Son-Starinets universal value $\hbar/(4 \pi k^{}_{B})$. 
\end{abstract}

\date{\today}

\pacs{03.75.Ss, 05.60.Gg, 51.20.+d, 05.30.Fk }

\maketitle


The unitary Fermi gas~(UFG) represents a dilute but strongly correlated system, where
the $s$-wave scattering between fermions saturates the unitarity bound for the  
cross section $\sigma(k) \leq 4\pi / k^2$ ($k$ being the relative wave vector of colliding particles).
The system is, therefore, characterized by the absence of intrinsic scales,
making it universal, i.e., independent of the details of the interaction.
On the other hand, the effects of interaction have to be treated nonperturbatively because of the lack of any small parameter.  
The extraordinary progress in experimental methods over the last decade has brought about the physical realization 
of such a system in the form of an ultracold gas of fermionic atoms~\cite{reviews}.
As a consequence, the UFG has provided a new paradigm for many strongly interacting Fermi systems,
attracting attention of theoretical physicists in various areas, including string theory, the quark-gluon plasma,
neutron stars, nuclei, and to a certain extent high-$T_c$ superconductivity~\cite{bcsbec}.

Over the last few years, an impressive effort has been underway, both experimentally and theoretically, to establish the physical 
properties of the UFG and reveal its strongly correlated nature. One of the most prominent manifestations of such strong 
correlations is the observation of nearly ideal hydrodynamic behavior~\cite{Turlapov,Cao1,Cao2}. Studies of the
transport properties of these systems are largely inspired by a conjecture formulated by 
Kovtun, Son, and Starinets (KSS) of the existence of a lower bound $\eta/s\geqslant\hbar/(4\pi k^{}_{B})$ 
on the ratio of the shear viscosity $\eta$ to the entropy density $s$ for any system~\cite{KSS}. As the bound is saturated 
for the case of strongly coupled ${\cal N}=4$ supersymmetric Yang-Mills 
theory, it is expected that strongly correlated quantum systems are close to this bound. 
Indeed, very different physical systems known to be strongly interacting appear to be very close to the KSS bound: 
i) the quark-gluon plasma created in heavy ion collisions at the RHIC obey $\eta/s\leqslant 0.4\hbar/k^{}_{B}$, 
ii) ultracold atomic gases at unitarity display $\eta/s\leqslant 0.5\hbar/k^{}_{B}$, see~\cite{SchaferTeaney} and references 
therein for an extensive overview. It has also been predicted that low-energy electrons in graphene monolayers are 
characterized by a low value of $\eta/s$, of the same order as that of the quark-gluon plasma and ultracold atomic gases~\cite{Mulleretal}.

In general, viscous (nonsuperfluid) hydrodynamics is characterized by two viscosity coefficients: the shear viscosity $\eta$ and 
the bulk viscosity $\zeta$. Contrary to the quark-gluon plasma, where the bulk viscosity is nonzero and 
can be a significant source of dissipation (especially near a phase transition), the bulk viscosity of the UFG vanishes as a 
result of scale invariance~\cite{Son,NishidaSon,TaylorRanderia}. The UFG is, therefore, an
excellent candidate for a perfect fluid, defined as the one with the lowest transport coefficients $\eta$ and $\zeta$ allowed by 
quantum mechanics. 

A large class of theoretical methods has been used to determine the transport coefficients of the UFG for homogeneous
and trapped systems \cite{BruunSmith,RupakSchafer,Schafer,Enssetal,Guoetal,Brabyetal,SalasnichToigo,LeClair,Mannarellietal}. 
Here, an {\it ab initio} calculation of the shear viscosity of the UFG is presented
within the framework of the Path Integral Monte Carlo (PIMC) approach~\cite{BDM}, which has been successfully used to 
compute other properties of the UFG~\cite{Bulgacetal,Magierskietal,Drutetal}.
Contrary to the previous {\it ab initio} calculations with the ``quenched" approximation, in which the fermion determinant is set to unity~\cite{NakamuraSakai,Meyer}, we compute the viscosity for a system with dynamical fermions.
The fact that such a fully dynamical calculation is at all possible is not \apriori obvious and should be regarded as one of our most important results.
While statistical errors are explicitly under control, we provide only a limited assessment of systematic effects 
(finite density and volume). From our results it is clear that those effects {\it can} be controlled.
While we focus our study on the 
shear viscosity, we have preliminary indications that the bulk viscosity vanishes at all temperatures, in agreement with the 
scale invariance arguments mentioned above. However, we defer more 
careful determinations of both viscosities as well as better control of systematic errors to future work.

Transport coefficients can be theoretically determined using linear response theory via the Kubo relations \cite{TaylorRanderia,Zubarev}. 
In order to apply such relations within the framework of PIMC calculations, we followed the method 
based on the stress-tensor correlators~\cite{NakamuraSakai,Teaney,Meyer}. Within this approach, the 
frequency-dependent shear viscosity is given by (in units such that $\hbar = k^{}_{B} = m = 1$)
\beq
\eta(\omega)=\pi\frac{\rho^{}_{xy,xy}(\bm{q}=0,\omega)}{\omega},
\eeq
while the static viscosity is defined in the limit of zero frequency: $\eta=\lim_{\omega\rightarrow 0^{+}}\eta(\omega)$. 
The spectral density $\rho^{}_{ij,kl}(\bm{q},\omega)$ is related to the 
imaginary-time (Euclidean) stress-tensor correlator $G^{}_{ij,kl}(\bm{q},\tau)$ by inversion of the relation
\beq
 G^{}_{i j, k l}(\bm{q},\tau)=\int_{0}^{\infty}\rho^{}_{i j, k l}(\bm{q},\omega)\,\dfrac{\cosh\left[ \omega(\tau-\beta/2)\right] }{\sinh\left( \omega\beta/2\right) }\,d\omega,
 \label{eqn:AnalitycalContinuation}
\eeq 
where $\beta=1/T$ is the inverse temperature. In turn, the stress-tensor correlator has the form
\beq
 G^{}_{i j, k l}(\bm{q},\tau)=\int d^{3}\bm{r}e^{-i\bm{q}\cdot\bm{r}}\avg{\hat{\Pi}^{}_{ij}(\bm{r},\tau)\hat{\Pi}^{}_{kl}(\bm{0},0)}, \label{eqn:StressTensorCorrelator}
\eeq
where the average is performed over the grand canonical ensemble, 
$\hat{O}(\tau)=e^{\tau(\hat{H}-\mu\hat{N})}\hat{O}e^{-\tau(\hat{H}-\mu\hat{N})}$,
$\hat{H}$ is the Hamiltonian of the system, $\mu$ is the chemical potential, and $\hat{N}$ is the particle number operator. 
The stress-tensor operator $\hat{\Pi}^{}_{ij}(\bm{r})$ is defined via the operator version of the Euler equation (summation over doubled index is assumed):
\beq
   i[\hat{j}^{}_{k}(\bm{r}),\hat{H}]=\partial^{}_{l}\hat{\Pi}^{}_{kl}(\bm{r}),
\eeq
where $\hat{j}^{}_{k}$ is the current operator. Since the current operator commutes neither with the kinetic-energy nor with the potential-energy parts
of the Hamiltonian, it is convenient to split the stress tensor into two parts: $\hat{\Pi}^{}_{kl}=\hat{\Pi}_{kl}^{(T)}+\hat{\Pi}_{kl}^{(V)}$. The kinetic-energy 
part $\hat{\Pi}_{kl}^{(T)}$ is well established and is the only contribution to the shear viscosity for a zero-range potential (see for 
example~\cite{Enssetal}). 
The potential-energy part $\hat{\Pi}_{kl}^{(V)}$ is more complicated, as defining the diagonal of the stress tensor is not trivial due to scale invariance, 
which is violated in our lattice calculations. Nevertheless, if we proceed with the stress tensor which on the lattice does not respect the sum rule 
$\int d^{3}\bm{r}\hat{\Pi}^{}_{ii}(\bm{r})=2\hat{H}$ imposed by the scale invariance~\cite{NishidaSon}, we obtain results consistent with $\zeta=0$. This matter is under further investigation.

Using the PIMC method, the stress-tensor correlator~(\ref{eqn:StressTensorCorrelator}) 
was evaluated at $\bm{q}=0$ for 51 points in imaginary time $\tau$, uniformly distributed in the interval $[0,\beta]$ on a 
spatial lattice of $8^3$ points. Increasing the number of $\tau$ points did not affect the final results.
A statistical ensemble of 5000 uncorrelated samples was generated at each temperature, thus reducing the statistical 
errors to a few percent (depending on the temperature and value of $\tau$).
To estimate the size of discretization errors, exploratory calculations on a $10^3$ lattice were performed. 
All the calculations presented here were performed with an average particle number density $n=N/V\approx 0.09$. 
The systematic errors associated with the stress-tensor correlator, related to finite volume effects as well as
effective-range corrections, are likely $\sim 10-15\%$~\cite{BDM, JED}. For a more detailed discussion, see Ref.~\cite{Supplemental}.

To determine $\eta$, one has to solve \eqn{eqn:AnalitycalContinuation} numerically, 
which is an ill-posed inversion problem, as there exist an infinite number of solutions that reproduce 
the correlator within its error bars. Therefore, estimating the shear viscosity requires
additional information. Besides the non-negativity of the viscosity $\eta(\omega)\geqslant 0$, 
the sum rule and the asymptotic tail behavior (see~\cite{TaylorRanderia} with subsequent corrections \cite{Enssetal,GoldbergerKhandker}) 
have been used as \apriori information. In the unitary limit these conditions read
\beq
\dfrac{1}{\pi}\int_{0}^{\infty}d\omega\left[ \eta(\omega) - \dfrac{C}{15\pi\sqrt{\omega}}\right] = \dfrac{\varepsilon}{3}, 
\eeq
where $C$ is Tan contact density~\cite{Tan} and $\varepsilon$ is the energy density. The energy density is obtained directly from PIMC 
calculations, while the contact density is taken from Ref.~\cite{Drutetal2}. 
Based on the results for the noninteracting Fermi gas, where $\eta^{}_{\textrm{FG}}(\omega)\propto\delta(\omega)$,
and those obtained within the $T$-matrix approach~\cite{Enssetal} or kinetic theory~\cite{Brabyetal}, the shear viscosity 
$\eta(\omega)$ is expected to be a continuous function with Gaussian-like structure at 
low frequencies, smoothly evolving into the asymptotic tail behavior $\eta(\omega\rightarrow\infty)\backsimeq\dfrac{C}{15\pi\sqrt{m\omega}}$. 
Moreover, we assume
that there is no sharp structure in the spectral density
in low frequency limit (associated, for example, with well defined quasiparticles), which could be overlooked 
during the inversion process.
We used these assumptions to construct the model used in the inversion procedure.

To perform the inversion we applied a methodology based on two complementary methods:
singular value decomposition (SVD) and maximum entropy method (MEM), both described in Ref.~\cite{MagierskiWlazlowski}. 
Since these methods are based on completely different approaches, a solution that is in agreement simultaneously with both 
of them is regarded as the most favorable scenario. In order to estimate the stability of the combined methods with respect 
to the algorithm parameters, the ``bootstrap" strategy was applied. Namely, about 200 reconstructions were performed,
with randomly generated initial parameters (within some reasonably chosen interval). 
The collected set of samples was subsequently used to evaluate the average value of the shear viscosity and the 
standard deviation (see~\cite{Supplemental} for details).

In \fig{fig:eta_per_n}, the dimensionless static shear viscosity $\eta/n$ is shown
as a function of $T/\varepsilon^{}_{F}$, where 
$\varepsilon^{}_{F}=(3\pi^{2}n)^{2/3}/2m$ is the Fermi energy of the noninteracting gas. 
The shear viscosity monotonically decreases with decreasing temperature. No drastic suppression of the viscosity 
below the critical temperature of the superfluid-normal phase transition $T_{c}^{}\simeq0.15\varepsilon^{}_{F}$ 
is observed. However, note that below $T_{c}^{}$ the coefficient $\eta$ describes 
the viscosity of the normal fluid component only.
The results on $8^{3}$ and $10^{3}$ lattices exhibit satisfactory agreement. 
Surprisingly, our results approach the predictions of kinetic theory already at $T\gtrsim0.3\varepsilon^{}_{F}$~\cite{BruunSmith}. 
Note that the PIMC results are significantly below all known results in the vicinity of $T_c^{}$.
\begin{figure}
\includegraphics[width=\pictsize\columnwidth]{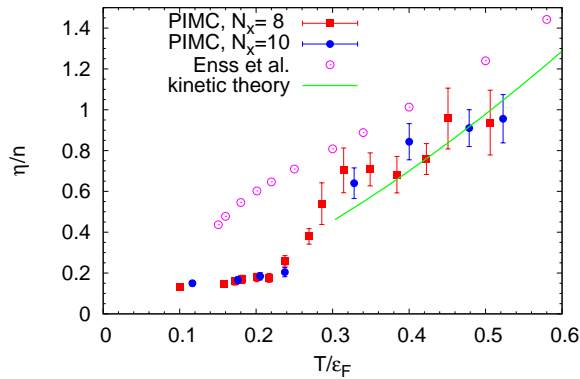}
\caption{ (Color online) The dimensionless static shear viscosity $\eta/n$ as a function of $T/\varepsilon^{}_{F}$ for an $8^3$ lattice 
(red) squares and $10^3$ lattice solid (blue) circles. The error bars only represent the stability of the combined (SVD and MEM) inversion procedure with 
respect to changes in the algorithm parameters. The (green) line depicts the prediction of kinetic theory~\cite{BruunSmith}.
For comparison, recent results of the $T$-matrix theory produced by Enss \etal are plotted as open (purple) circles~\cite{Enssetal}.
\label{fig:eta_per_n} }
\end{figure}

\begin{figure}[t]
\includegraphics[width=\pictsize\columnwidth]{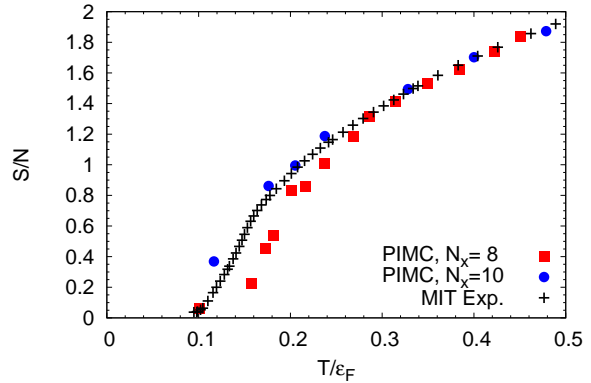}
\caption{(Color online) Entropy per particle as a function of $T/\varepsilon^{}_{F}$ for the $8^3$ lattice in (red) squares and $10^3$ lattice in (blue) circles. The entropy per particle extracted from the recent MIT measurement~\cite{MIT_exp} is plotted with (black) crosses.
\label{fig:entropy} }
\end{figure}

In~\fig{fig:entropy}, the value of the entropy obtained from PIMC calculations is shown (extracted as in Ref.~\cite{BDM}), together with the results extracted from the recent high-precision MIT measurement~\cite{MIT_exp}. 
For temperatures $T>0.25\varepsilon^{}_{F}$, both lattices reproduce experimental data reasonably well. At low temperatures 
$T<0.25\varepsilon^{}_{F}$ the $8^3$-lattice results deviate from the measurements, producing systematically lower values. 
On the other hand, the $10^3$-lattice results reproduce correctly the temperature dependence of the entropy, yet slightly overestimating 
the experimental values. These discrepancies are attributed to systematic errors that are known 
to be present at low temperatures even for larger lattices~\cite{Drutetal}.
Consequently, we expect the ratio $\eta/s$ to be significantly affected by uncertainties related to the entropy at low temperatures.

\begin{figure}[t]
\includegraphics[width=\pictsize\columnwidth]{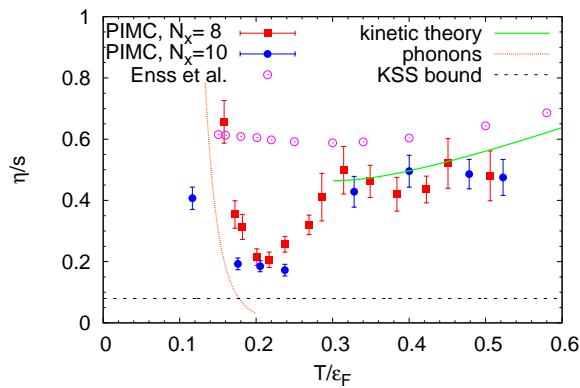}
\caption{ (Color online) Ratio of the shear viscosity to entropy density $\eta/s$ as a function of $T/\varepsilon^{}_{F}$ for an $8^3$ lattice (red) 
squares and $10^3$ lattice (blue) circles.
The error bars only represent the stability of the combined (SVD and MEM) inversion procedure with respect to the change of algorithm parameters 
and do not include systematic errors of the entropy determination.
Results of the $T$-matrix theory are plotted by open (purple) circles~\cite{Enssetal}. In the high- and low-temperature regimes, known asymptotics are 
depicted: for $T>0.3\varepsilon^{}_{F}$ the prediction of kinetic theory~\cite{BruunSmith} as a green line, and for $T<0.2\varepsilon^{}_{F}$ the contribution from 
phonon excitations~\cite{RupakSchafer} as a brown line. The KSS bound appears as a dashed black line.
\label{fig:eta_per_s} }
\end{figure}

In \fig{fig:eta_per_s} the ratio $\eta/s$ is presented as a function of temperature. The PIMC calculations reveal the existence of a deep 
and rather narrow minimum in $\eta/s$ at temperatures around $0.20-0.25\varepsilon^{}_{F}$, which is above $T_{c}^{}$. 
Again, the ratio $\eta/s$ is located around the kinetic theory predictions already at $T\gtrsim0.3\varepsilon^{}_{F}$~\cite{BruunSmith}.
The estimation of the $\eta/s$-ratio reveals $(\eta/s)^{}_{\textrm{min}}\approx0.2$ as the most probable value for the minimum.
This result is about 2.5 times higher than the KSS bound $\eta/s\geqslant 1/4\pi\approx 0.08$. Such a low value has been reported only for 
pure gluons as a result of lattice calculations~\cite{NakamuraSakai, Meyer}.

The minimum value for the ratio $(\eta/s)^{}_{\textrm{min}}\approx0.2$, is significantly lower than predictions 
of all current calculations, which yield a minimum $\simeq 0.5$. However, these methods are 
in principle unreliable when applied to the UFG at $T\simeq T_c^{}$, where 
the minimum appears. Moreover, the $\eta/s$ ratio calculated from PIMC simulations
is also significantly lower than the experimental measurements~\cite{Turlapov,Cao1,Cao2}, which also give the value $\simeq 0.5$. 
Note, however that these measurements are performed in trapped systems. The trap-averaged viscosity $\avg{\eta/n}=\frac{1}{N\hbar}\int \eta(\bm{r})\,d^{3}\bm{r}$ may affect the determination of the minimum value. 
To solve this puzzle, one should apply an averaging procedure to the uniform case results, 
using, e.g., local density approximation.
It is well known that this procedure leads to a divergence due to the violation of the hydrodynamic description 
at the edges of the cloud~\cite{SchaferChafin}. To perform a reliable averaging procedure the collisionless edges 
should be treated using kinetic theory. 
This, however, is a hard task that requires the knowledge of second-order transport coefficients like the relaxation time, 
which are currently poorly known.

Since our main result for the minimal value of $\eta/s$ is significantly lower than other predictions as well as experimental results, we have performed 
exploratory calculations to estimate the size of systematic effects. We have checked the stability of the inversion procedure with respect to the 
default model as well the impact of the nonzero value of the effective range, see~\cite{Supplemental} for details. 
Our conservative estimation indicates that the minimal value of the $\eta/s$-ratio is lower than $0.45$. 

In summary, we have presented an attempt to determine the shear viscosity of the UFG through an
{\it ab initio} PIMC approach.
The minimum value of the $\eta/s$ ratio was estimated to be lower than $0.45$ with the most probable value being 
$(\eta/s)^{}_{\textrm{min}}\approx 0.2$, located around $T\approx0.20-0.25\varepsilon^{}_{F}$. 
This value is close to the KSS bound and suggests that the unitary Fermi gas is the best candidate for the 
perfect fluid. As our results can be significantly affected by systematic errors, further and more precise investigations are called for.

We thank M. Zwierlein for making the MIT experiment data~\cite{MIT_exp} available to us
and T. Enss for providing us with the $T$-matrix results~\cite{Enssetal}.
We are indebted to A.~Bulgac for instructive discussions and a careful reading of the manuscript.
We acknowledge support under U.S. DOE Grant No. DE-FC02-07ER41457, and
Contract No. N N202 128439 of the Polish Ministry of Science.
One of the authors (G.W.) acknowledges the Polish Ministry of Science for
the support within the program ``Mobility Plus \!-\! I edition'' under Contract No. 628/MOB/2011/0.
Calculations reported here have been, in part, performed at the Interdisciplinary Centre for Mathematical and Computational Modelling (ICM) at Warsaw University and on the University of Washington Hyak cluster funded by the NSF MRI Grant No. PHY-0922770.


\newpage
\begin{center}
{\bf Supplemental online material for:}\\
{\bf ``Shear Viscosity of a Unitary Fermi Gas''}\\
\end{center}

\begin{small}
\noindent
This supplemental material provides the details concerning
the inversion procedure, and the discussion of systematic errors.
\end{small}
\setcounter{equation}{0}
\setcounter{figure}{0}

\section{Analytic continuation \& inversion}

The calculation of dynamic response functions, such as susceptibilities, viscosities and conductivities, entails computing 
real-time correlation functions. Since PIMC calculations are performed in imaginary time, one faces the problem of analytic 
continuation. As real-time and imaginary-time correlations share the same spectral density $\rho$, this problem can be recast 
as an inversion problem in which one attempts to find $\rho$ with imaginary-time correlation functions as the starting point.
In the case of the shear viscosity this inversion problem is given by the equation (for zero momentum $\bm{q}=0$):
\beq
 G^{}_{xy,xy}(\tau^{}_{i})=\int_{0}^{\infty}\rho^{}_{xy,xy}(\omega)\,K(\omega,\tau^{}_{i})\,d\omega,
 \label{eqn:AnalitycalContinuation}
\eeq
with the kernel
\beq
 K(\omega,\tau)=\dfrac{\cosh\left( \omega(\tau-\beta/2)\right) }{\sinh\left( \omega\beta/2\right) }.
\eeq
The correlator $G^{}_{xy,xy}(\tau^{}_{i})$ is determined within PIMC
with certain accuracy
$\Delta G^{}_{xy,xy}(\tau^{}_{i})$ for a finite set of points $\tau^{}_{1},\tau^{}_{2},\ldots,\tau^{}_{N}$ 
uniformly distributed within interval $[0,\beta]$ (we used $N=51$ for $8^3$ and $N=41$ for $10^3$ lattice).
For brevity we skip subscripts ($xy,xy$) hereafter. 
By definition, the spectral density vanishes at zero frequency, while the kernel has a pole 
$K(0,\tau)\rightarrow \infty$. Consequently, the expression $\rho(0)K(0,\tau)$ is not a well defined quantity
for the lower limit of the integral~(\ref{eqn:AnalitycalContinuation}).
However, since we are interested in the static shear viscosity
\beq
 \eta=\lim_{\omega\rightarrow 0^{+}}\eta(\omega)=\lim_{\omega\rightarrow 0^{+}}\dfrac{\pi\rho(\omega)}{\omega}
\eeq
we can formulate the problem~(\ref{eqn:AnalitycalContinuation}) as
\beq
 G(\tau^{}_{i})=\dfrac{1}{\pi}\int_{0}^{\infty}\eta(\omega)\,\tilde{K}(\omega,\tau^{}_{i})\,d\omega,
\eeq
with a new kernel $\tilde{K}(\omega,\tau)=\omega\,K(\omega,\tau)$. Note that the new kernel is well defined 
at zero frequency $\tilde{K}(0,\tau)=2/\beta$. Moreover, the frequency dependent shear viscosity
can be now directly determined, without taking the limit $\lim_{\omega\rightarrow 0^{+}}\rho(\omega)/\omega$ 
which could be difficult to realize numerically.

From the symmetry of the kernel we infer a simple relation for the correlator $G(\tau)=G(\beta-\tau)$, which is
simply a result of the bosonic character of the stress tensor, and is fulfilled in our calculations within error bars. 
Consequently, we may restrict the inversion to the interval $\tau\in[0,\beta/2]$.
Moreover, we have replaced $G(\tau)\leftarrow[G(\tau)+G(\beta-\tau)]/2$ in order to obtain 
``smoother'', symmetric correlator. The problem (\ref{eqn:AnalitycalContinuation}) 
is supplemented with external constraints: non-negativity of the shear viscosity, sum rule and asymptotic tail behavior.
From the known tail behavior we obtain that the correlator $G(\tau)$ should have poles for $\tau\in\{0,\beta\}$. 
Indeed, the contribution from the tail for $\tau=0$ reads (in units such that $\hbar = k_{B}^{} = m = 1$)
\beq
 \dfrac{1}{\pi}\int_{\omega^{}_{\textrm{large}}}^{+\infty}\dfrac{C}{15\pi\sqrt{\omega}}\,\omega\,d\omega=+\infty,
 \label{eqn:tailcontrib}
\eeq
where $\omega^{}_{\textrm{large}}\gg 1$ is assumed to be large enough to approximate 
the viscosity by the tail behavior and $K(\omega,0)$ by unity. The above result indicates that the
the correlator at small~$\tau$ does not carry much information about the low-frequency part of the shear viscosity.
The most important information about the shear viscosity at low frequencies is encoded in the region located 
around $\tau=\beta/2$.
Moreover, the PIMC approach does not provide us with a correlator $G(\tau)$ that acquires 
extremely high values at the edges of domain. We attribute this anomaly to the systematic error related to the 
existence of the cut-off in the momentum space, which implies a cut-off in the frequencies  
at $\omega^{}_{\textrm{max}}\sim \pcut^{2}/2$. Since for small $\tau$, $G(\tau)$ 
is strongly affected by systematic errors and does not provide significant information for estimation 
of the static shear viscosity, we remove from the analysis few first points of $G(\tau^{}_{i})$ (typically about 3).

The inversion was performed using a combination of two complementary methods: 
Singular Value Decomposition (SVD) and Maximum Entropy Method (MEM).
We have applied a variant of the MEM referred to as \textit{self-consistent MEM} in the paper~\cite{MagierskiWlazlowskiSupp}.
Within this method, the \apriori model solution is not fixed (as it is in standard MEM) but evolves 
simultaneously with the solution within the space of functions spanned by admissible models. 

\begin{figure}
\includegraphics[width=\pictsize\columnwidth]{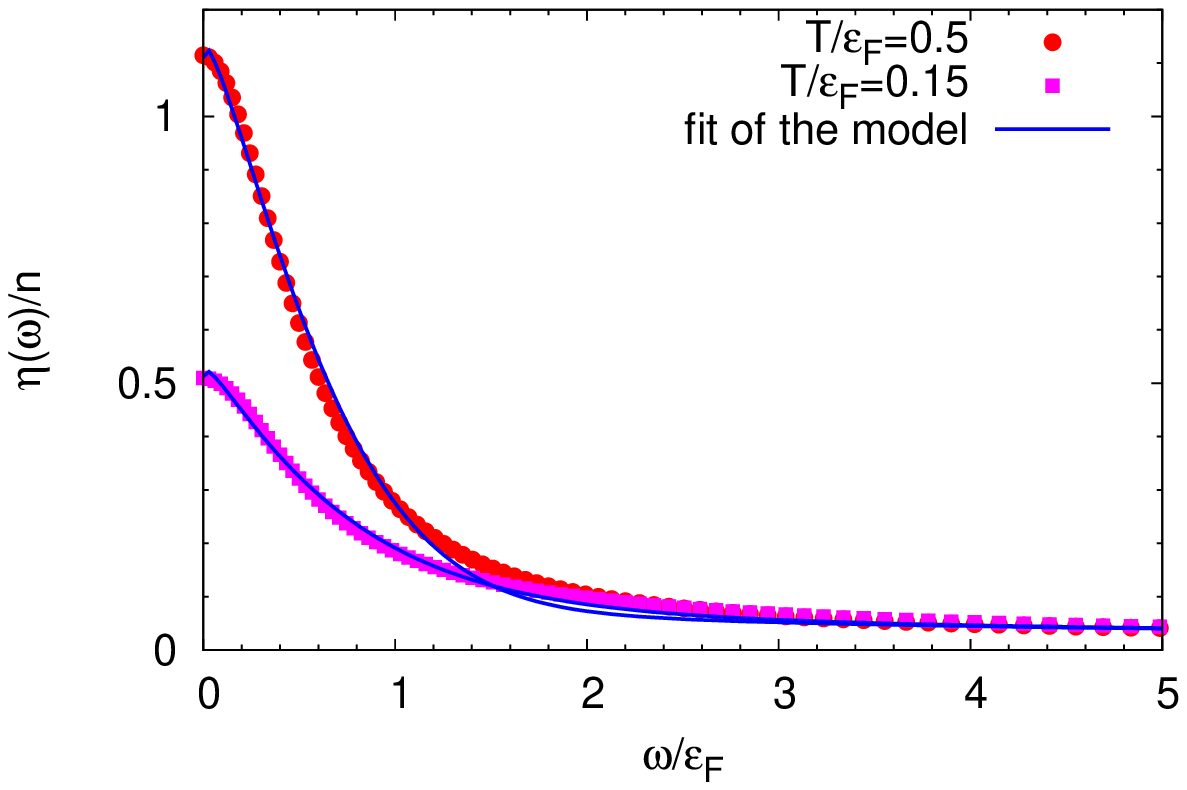}
\caption{ (Color online) Fit of the proposed model (blue line) to the results produced by T-matrix theory (red and purple symbols)~\cite{EnssetalSupp} for two selected temperatures.  
\label{fig:fitmodelEnss} }
\end{figure}

The class of models for the self-consistent MEM is defined as
\bea
M(\omega,\{m,\sigma,c,\alpha^{}_{1},\alpha^{}_{2}\}) = f(\omega,\{\alpha^{}_{1},\alpha^{}_{2}\})\dfrac{C}{15\pi\sqrt{\omega}}\nonumber \\
 +[1\!-\!f(\omega,\{\alpha^{}_{1},\alpha^{}_{2}\})] N(\omega,\{m,\sigma,c\}),
\eea
where
\beq
 N(\omega,\{m,\sigma,c\})=\dfrac{c}{\sqrt{2\pi\sigma^{2}}}\exp\left( -\dfrac{(\omega-m)^{2}}{2\sigma^{2}}\right), 
\eeq
and
\beq
 f(\omega,\{\alpha^{}_{1},\alpha^{}_{2}\})=e^{-\alpha^{}_{1}\alpha^{}_{2}}\dfrac{e^{\alpha^{}_{1}\omega}-1}{1+e^{\alpha^{}_{1}(\omega-\alpha^{}_{2})}},
\eeq 
such that $f(\omega\rightarrow 0)\rightarrow 0$ and $f(\omega\rightarrow \infty)\rightarrow 1$. The function $f(\omega,\{\alpha^{}_{1},\alpha^{}_{2}\})$ 
guarantees a smooth change of behavior between Gaussian dependence and the known tail dependence.
The set of five parameters $\{m,\sigma,c,\alpha^{}_{1},\alpha^{}_{2}\}$ describes the available degrees of freedom of the model for the self-consistent 
MEM. To initialize the model for the first iteration we fit it to the SVD solution. 
We have found that the proposed model can reasonably describe the SVD solution for all considered cases. 
We have also checked that the solutions provided by the T-matrix theory~\cite{EnssetalSupp} can be well reproduced by the used model, 
see~\fig{fig:fitmodelEnss}. We also applied the T-matrix results, to estimate the range of frequencies $\omega^{}_{\textrm{tail}}$ 
above which the solution is well reproduced by the tail behavior.

The uncertainty related to the inversion procedure was estimated using the ``bootstrap" strategy.
The bootstrap sample consists of about 200 launches of the algorithm with randomly generated (from some reasonable intervals) 
parameters: i) $\omega^{}_{max}$ - the upper limit of integration, ii) $\omega^{}_{\textrm{tail}}$ - the point where the universal tail behavior 
starts, iii) $\alpha$ - the parameter for the MEM algorithm. To represent the MEM solution $\eta(\omega)$, we used mesh of $100$ points uniformly distributed within range $\omega\in[0,\omega^{}_{max}]$.
The static shear viscosity $\eta(0)$ is computed as the average over the 
bootstrap collection, while the uncertainty is determined by the standard deviation. 

\begin{figure}
\includegraphics[width=\pictsize\columnwidth]{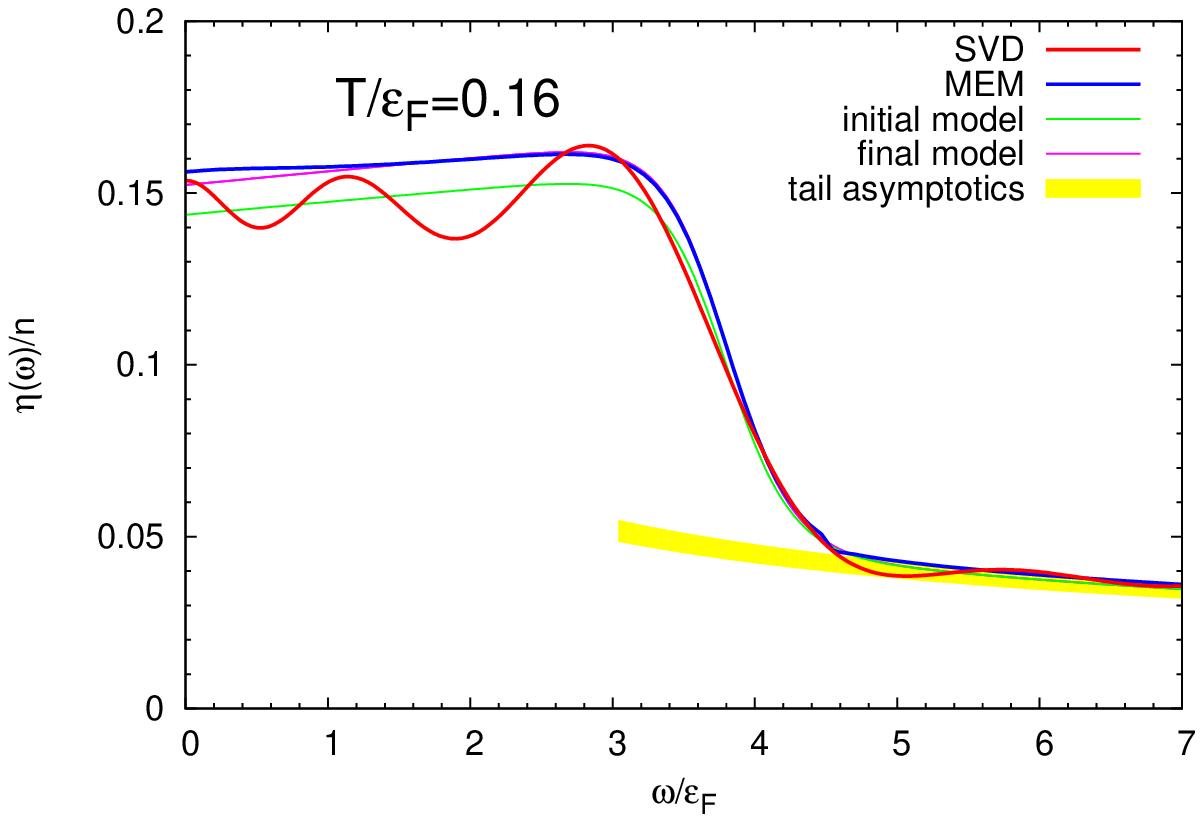}
\includegraphics[width=\pictsize\columnwidth]{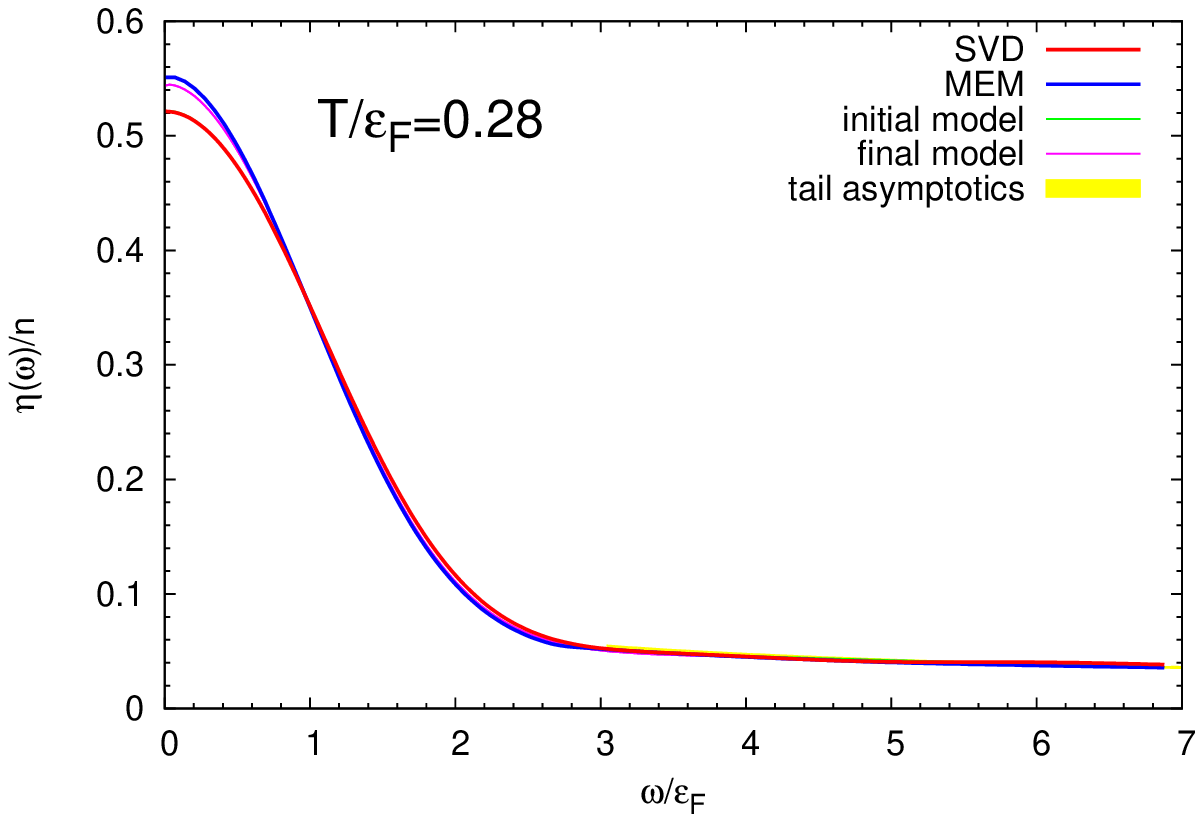}
\includegraphics[width=\pictsize\columnwidth]{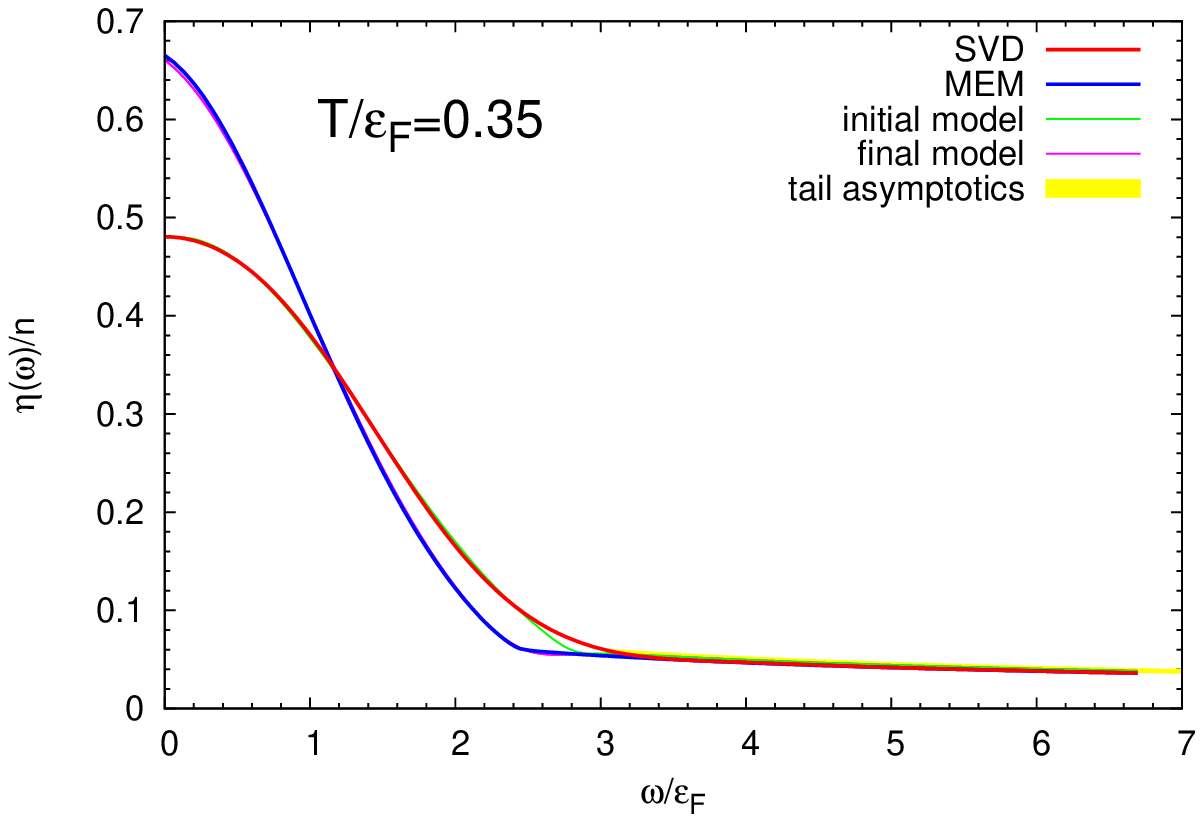}
\caption{ (Color online) The frequency dependent shear viscosity in units of density for three different temperatures obtained from the Quantum Monte Carlo calculations with $8^3$-lattice. The SVD solution is used to create the initial model for the self-consistent MEM.
\label{fig:MCeta} }
\end{figure}
In~\fig{fig:MCeta} we present a sample of results for three temperatures. For temperatures above $0.25\varepsilon^{}_{F}$ we observe 
that $\eta(\omega)$ possesses a Gaussian-like structure at low frequencies. This structure is smeared out at low temperatures.  Note also that for 
low temperatures both the SVD and MEM solutions are similar, while at larger temperatures the SVD method produces solutions of lower static 
shear viscosity $\eta(0)$ than the MEM approach. Such discrepancy is permissible, since the SVD method gives only the projection of the solution 
onto the relatively small subspace where the inverse problem is well defined~\cite{MagierskiWlazlowskiSupp}.

\section{Discussion of systematic errors}
In this section we present exploratory estimations of corrections arising from systematic errors. Since the results for two different lattice sizes 
$N^{}_{x}=8$ and $10$ agree sufficiently well, we shall focus on the corrections attributed to the inversion procedure and effective range. 

\begin{figure}[t]
\includegraphics[width=\pictsize\columnwidth]{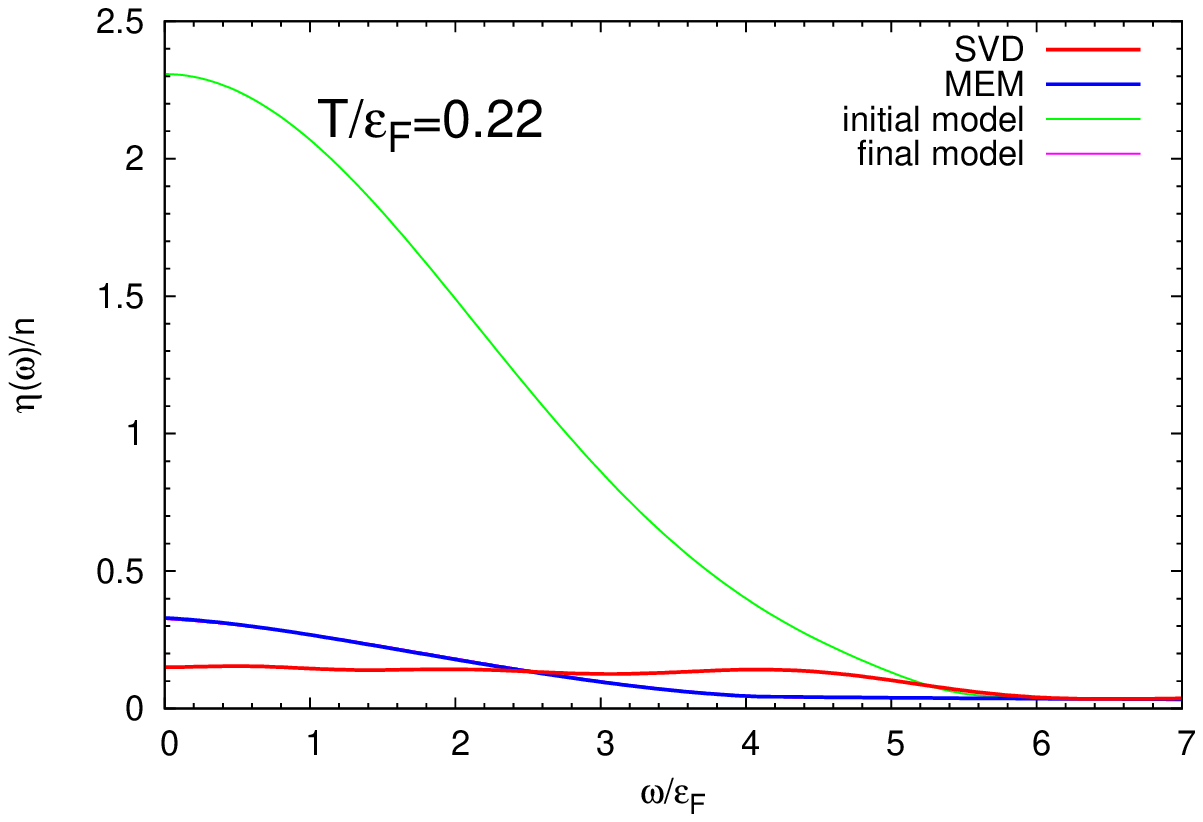}
\caption{ (Color online) The frequency dependent shear viscosity in units of density obtained from the self-consistent MEM when initialized with exaggerated model.
\label{fig:MCetau_T020} }
\end{figure}
\begin{figure}[t]
\includegraphics[width=\pictsize\columnwidth]{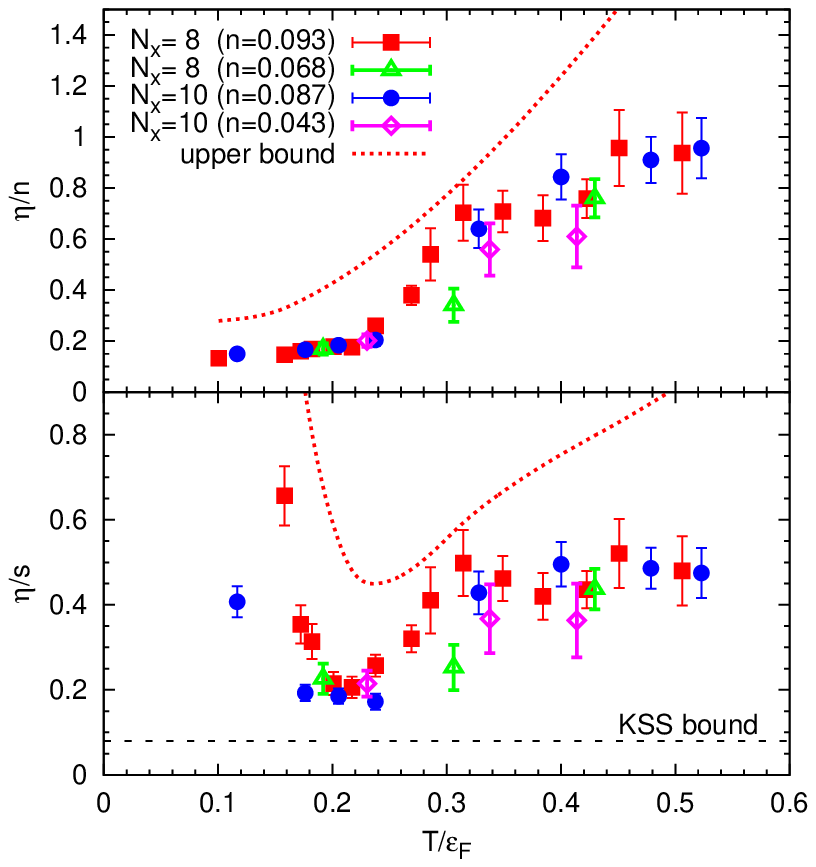}
\caption{ (Color online) The dimensionless static shear viscosity $\eta/n$ (upper panel) and the ratio of the shear viscosity to the entropy density 
$\eta/s$ (lower panel) as a function of $T/\varepsilon^{}_{F}$. Filled (red) squares and filled (blue) circles show the results for the $8^3$ and $10^3$ 
lattices, respectively, with $n\approx0.09$ as presented in the main paper. Results for reduced densities are shown as open (green) triangles and 
open (purple) diamonds. The (red) dotted line shows a conservative estimate for the upper bound. For the $\eta/s$ ratio, the upper bound 
includes also possible corrections arising from systematic errors in the estimation of the entropy. The dashed (black) line shows the KSS bound.
\label{fig:sup_eta_per_n} }
\end{figure}

According to our experience the methodology based on the combination of the SVD method and the self-consistent MEM provides the 
most reliable results. However one may ask the following question: What is the maximal value of the shear viscosity 
which is simultaneously consistent with QMC data and with the external constraints? In order to answer this question 
we have applied the self-consistent MEM with the initial model predicting a significantly higher value of the viscosity 
than all known predictions for the studied temperatures, see~\fig{fig:MCetau_T020}. By construction, the initial 
model does not satisfy the sum rule. Since we start with this exaggerated model in each subsequent iteration the self-consistent MEM 
systematically reduces $\eta(\omega)$ and thus provides us with an upper bound for the viscosity. In general, we observe 
that the produced solutions are in clear disagreement with the solutions provided by the SVD method. 
The estimated upper bound for $\eta/s$ reveals the value $(\eta/s)^{}_{\textrm{min}}\lesssim0.45$ (see \fig{fig:sup_eta_per_n}), where we used the smallest 
obtained value for the entropy density. Namely, we used $8^3$-lattice results, which clearly are affected by systematic errors, especially in the regime 
where the minimum is located (see Fig.~2 in the main paper), which artificially enhances $\eta/s$.
It should also be noted that the generated values for the static viscosity $\eta$ significantly overestimate all known results for uniform 
systems for temperatures $T>0.4\varepsilon^{}_{F}$. As a consequence, it is difficult to obtain a smooth connection 
of the upper limit predictions with the known behavior of the viscosity at high temperatures 
$\eta/n\approx 2.77(T/\varepsilon^{}_{F})^{3/2}$~\cite{EnssetalSupp}. This problem is absent for the most probable solutions, where the 
static shear viscosity approaches smoothly results of kinetic theory. 

To obtain stable results for the inversion procedure with respect to change of the algorithm parameters it is necessary to include external constraints, especially the sum rule together with the asymptotic tail behavior. Both of these constraints as ingredient include Tan's contact density. 
Since we were not able to determine the value of the contact  within present work from the asymptotics of the momentum distribution (which requires very low densities), we used values obtained in earlier work~\cite{Drutetal2Supp}. This may introduce additional systematic error into our calculations, but we have checked that variation 
of the contact by $\pm 10\%$ generates change of the shear viscosity within plotted error bars.

Recent results~\cite{CarlsonetalSupp,DrutetalSupp,PriviteraCaponeSupp,JEDSupp} suggest that we should expect significant modifications arising from effective range 
corrections as the product $r^{}_{\text{eff}}k^{}_{F}\approx 0.6$ remains non-negligible. To check the impact of these corrections we performed 
exploratory calculations with reduced density. For the $8^3$ lattice we reduced the density to $n\approx0.07$ (we decided not to reduce it 
further to avoid shell effects) and for the $10^3$ lattice to $n\approx0.04$ which corresponds to $r^{}_{\text{eff}}k^{}_{F}\approx 0.5$ and $0.4$, 
respectively. Performing calculations at significantly lower densities, while keeping $T/\epsilon_F^{}$ sufficiently low, is an extremely 
time-consuming task. We therefore decided to perform calculations only for a few selected temperatures. Our limited number 
of tests indicate that decreasing finite-range effects yields slightly lower viscosities for temperatures $T/\varepsilon^{}_{F}>0.3$, 
see \fig{fig:sup_eta_per_n}. In the regime of temperatures where the minimum of $\eta/s$-ratio is located, all the results agree within the 
inversion error bars. Hence, we conclude that the estimated upper bound for the shear viscosity is rather conservative. All the theories that 
produce viscosities above our estimations of the upper bound are in clear disagreement with our results.


\begin{thebibliography}{99}
\bibitem{reviews} 
	S Giorgini, L.P. Pitaevskii, S. Stringari, 
	Rev. Mod. Phys.  {\bf 80}, 1215 (2008); 
        I. Bloch, J. Dalibard, W. Zwerger, 
        Rev. Mod. Phys. {\bf 80}, 885 (2008).

\bibitem{bcsbec} 
	\textit{The BCS-BEC crossover and the unitary Fermi Gas} 
		Lecture Notes in Physics, edited by W. Zwerger (Springer-Verlag, Berlin, 2012), Vol. 836.    

\bibitem{Turlapov} 
	A. Turlapov, J. Kinast, B. Clancy, L. Luo, J. Joseph, J.E. Thomas, 
	J. Low Temp. Phys. {\bf 150}, 567 (2008).

\bibitem{Cao1} 
	C. Cao, E. Elliott, J. Joseph, H. Wu, J. Petricka, T. Schaefer, J. E. Thomas, 
	Science {\bf 331}, 58 (2011).

\bibitem{Cao2} 
	C. Cao, E. Elliott, H. Wu and J.E. Thomas, 
	New J. Phys. {\bf 13}, 075007 (2011).

\bibitem{KSS} 
	P.K. Kovtun, D.T. Son, A.O. Starinets, 
	Phys. Rev. Lett. {\bf 94}, 111601 (2005).

\bibitem{SchaferTeaney} 
	T. Sch\"{a}fer and D. Teaney, 
	Rep. Prog. Phys. {\bf 72}, 126001 (2009).

\bibitem{Mulleretal} 
	M. M\"{u}ller, J. Schmalian, and L. Fritz, 
	Phys. Rev. Lett. {\bf 103}, 025301 (2009).

\bibitem{Son} 
	D.T. Son, 
	Phys. Rev. Lett. {\bf 98}, 020604 (2007).

\bibitem{NishidaSon} 
	Y. Nishida and D.T. Son, 
	Phys. Rev. D {\bf 76}, 086004 (2007).

\bibitem{TaylorRanderia} 
	E. Taylor and M. Randeria, 
	Phys. Rev. A {\bf 81}, 053610 (2010).
	
\bibitem{BruunSmith} 
	G.M. Bruun and H. Smith, 
	Phys. Rev. A {\bf 72}, 043605 (2005); 
	Phys. Rev. A {\bf 75}, 043612 (2007).

\bibitem{RupakSchafer} 
	G. Rupak and T. Sch\"{a}fer, 
	Phys. Rev. A {\bf 76}, 053607 (2007).

\bibitem{Schafer} 
	T. Sch\"{a}fer, 
	Phys. Rev. A {\bf 76}, 063618 (2007).

\bibitem{Enssetal} 
	T. Enss, R. Haussmann, W. Zwerger, 
	Ann. Phys. {\bf 326}, 770 (2011).

\bibitem{Guoetal} 
	H. Guo, D. Wulin, C.-C. Chien, and K. Levin, 
	Phys. Rev. Lett. {\bf 107}, 020403 (2011).

\bibitem{Brabyetal} 
	M. Braby, J. Chao and T. Sch\"{a}fer, 
	New J. Phys. {\bf 13}, 035014 (2011).

\bibitem{SalasnichToigo} 
	L. Salasnich, F. Toigo, 
	J. Low Temp. Phys. {\bf 165}, 239 (2011).

\bibitem{LeClair} 
	A. LeClair, 
	New J. Phys. {\bf 13}, 055015 (2011).

\bibitem{Mannarellietal}
	M.~Mannarelli, C.~Manuel, L.~Tolos, arXiv:1201.4006v1.

\bibitem{BDM} 
	A. Bulgac, J.E. Drut, P. Magierski, 
	Phys. Rev. A {\bf 78} 023625 (2008).
	
\bibitem{Bulgacetal} 
	A. Bulgac, J.E. Drut, and P. Magierski, 
	Phys. Rev. Lett. {\bf 96} 090404 (2006).

\bibitem{Magierskietal} 
	P. Magierski, G. Wlaz\l{}owski, A. Bulgac and J.E. Drut, 
	Phys. Rev. Lett. {\bf 103}, 210403 (2009);
        P. Magierski, G. Wlaz\l{}owski, and A. Bulgac, 
        Phys. Rev. Lett. {\bf 107}, 145304 (2011).

\bibitem{Drutetal} 
	J.E. Drut, T.A. L\"{a}hde, G. Wlaz\l{}owski, P. Magierski, 
	Phys. Rev. A {\bf 85}, 051601(R) (2012).

\bibitem{NakamuraSakai} 
	A. Nakamura, S. Sakai, 
	Phys. Rev. Lett. {\bf 94}, 072305 (2005).
	
\bibitem{Meyer} 
	H.B. Meyer, 
	Phys. Rev. D {\bf 76}, 101701(R) (2007); 
	Phys. Rev. Lett. {\bf 100}, 162001 (2008).
	
\bibitem{Zubarev} 
	D.N. Zubarev, 
	\textit{Nonequilibrium Statistical Thermodynamics}, (Consultants Bureau, New York, 1974).

\bibitem{Teaney} 
	D. Teaney, 
	Phys. Rev. D {\bf 74}, 045025 (2006).
	
\bibitem{JED}
	J.E. Drut,
	Phys. Rev. A {\bf 86}, 013604 (2012).

\bibitem{Supplemental} See Supplemental Material for technical details concerning the inversion procedure and the discussion of systematic errors.

\bibitem{GoldbergerKhandker} 
	W.D. Goldberger and Z.U. Khandker, 
	Phys. Rev. A {\bf 85}, 013624 (2012).

\bibitem{Tan} 
	S. Tan, Ann. 
	Phys. {\bf 323}, 2952 (2008).

\bibitem{Drutetal2} 
	J.E. Drut, T.A. L\"{a}hde, T. Ten, 
	Phys. Rev. Lett. {\bf 106}, 205302 (2011).

\bibitem{MagierskiWlazlowski} 
	P. Magierski, G. Wlaz\l{}owski, 
	Comput. Phys. Commun. {\bf 183}, 2264 (2012).

\bibitem{MIT_exp}
	M.J.H.~Ku, A.T.~Sommer, L.W.~Cheuk, M.W.~Zwierlein,
	Science {\bf 335}, 563 (2012).

\bibitem{SchaferChafin}
	T. Sch\"{a}fer and C. Chafin, 
	\textit{Scaling Flows and Dissipation in the Dilute Fermi Gas at Unitarity} Chap. 10
	in \textit{BCS-BEC Crossover and the Unitary Fermi Gas}, edited by W. Zwerger (Springer, Berlin, 2012).

\end{thebibliography}

\begin{thebibliography}{99}

\bibitem{MagierskiWlazlowskiSupp} 
	P. Magierski, G. Wlaz\l{}owski, 
	Comput. Phys. Commun. {\bf 183}, 2264 (2012).

\bibitem{EnssetalSupp} 			
	T. Enss, R. Haussmann, W. Zwerger, 
	Ann. Phys. {\bf 326}, 770 (2011).
		
\bibitem{Drutetal2Supp} 
	J.E. Drut, T.A. L\"{a}hde, T. Ten, 
	Phys. Rev. Lett. {\bf 106}, 205302 (2011).
	
\bibitem{CarlsonetalSupp}
	J. Carlson {\it et al.}, 
	Phys. Rev. A {\bf 84}, 061602(R) (2011).

\bibitem{PriviteraCaponeSupp}		
	A. Privitera and M. Capone,
	Phys. Rev. A {\bf 85}, 013640 (2012).
	
\bibitem{JEDSupp}				
	J.E. Drut,
	Phys. Rev. A {\bf 86}, 013604 (2012).
	
\bibitem{DrutetalSupp} 			
	J.E. Drut, T.A. L\"{a}hde, G. Wlaz\l{}owski, P. Magierski, 
	Phys. Rev. A {\bf 85}, 051601(R) (2012).
	
\end{thebibliography}
\end{document}